\documentclass[aps,amsmath,amssymb,nofootinbib,superscriptaddress,showpacs,floatfix,prl,twocolumn]{revtex4-1}

\usepackage{latexsym}
\usepackage{graphicx}
\usepackage{times,psfrag,subfigure}
\usepackage{enumerate}
\usepackage{amsmath}
\usepackage{dsfont}
\usepackage{dcolumn}
\usepackage{bm,bbm}
\usepackage{color}
\usepackage{latexsym,amsmath,amssymb,bm,euscript}
\usepackage{dsfont}
\usepackage{textcomp}
\usepackage{tabularx}
\usepackage{setspace}
\usepackage{ctable}
\usepackage{sidecap}
\usepackage{placeins}
\usepackage{threeparttable}
\usepackage{multirow}

\def\Z{\mathds{Z}}
\def\R{{R}}
\def\T{\mathcal{T}}
\def\C{{C}}
\def\S{\mathcal{S}}
\def\d{{\rm d}}
\def\H{\mathcal{H}}
\def\K{\mathcal{K}}

\newcommand{\up}{\uparrow}
\newcommand{\sech}{\text{sech}~}
\newcommand{\dn}{\downarrow}
\newcommand{\beq}{\begin{equation}}
\newcommand{\eeq}{\end{equation}}
\newcommand{\beqarray}{\begin{eqnarray}}
\newcommand{\eeqarray}{\end{eqnarray}}

\begin{document}

\allowdisplaybreaks

\title{Dimensional hierarchy of fermionic interacting topological phases}

\date{\today}

\author{Raquel Queiroz}
\email{r.queiroz@fkf.mpg.de}
\affiliation{Max-Planck-Institut f\"ur Festk\"orperforschung, Heisenbergstrasse 1, D-70569 Stuttgart, Germany}
\author{Eslam Khalaf}
\email{e.khalaf@fkf.mpg.de}
\affiliation{Max-Planck-Institut f\"ur Festk\"orperforschung, Heisenbergstrasse 1, D-70569 Stuttgart, Germany}
\author{Ady Stern}
\affiliation{Department of Condensed Matter Physics, Weizmann Institute of Science, Rehovot 76100, Israel}

\begin{abstract}
We present a dimensional reduction argument to derive the classification reduction of fermionic symmetry protected topological phases in the presence of interactions.
The dimensional reduction proceeds by relating the topological character of a $d$-dimensional system to the number of zero-energy bound states localized at zero-dimensional topological defects present at its surface. This correspondence leads to a general condition for symmetry preserving interactions that render the system topologically trivial, and allows us to explicitly write a quartic interaction to this end. Our reduction shows that all phases with topological invariant smaller than $n$ are topologically distinct, thereby reducing the non-interacting $\Z$ classification to $\Z_n$.
\end{abstract}

\date{\today}


\maketitle

Topological phases of matter are presently one of the main research topics in condensed matter physics. The discovery of time-reversal invariant topological insulators (TIs) \cite{Konig2012,Hsieh2008}  and superconductors (TSCs) has led, among other results, to a systematic classification of topological phases of non-interacting fermions in a general spatial dimension and symmetry class \cite{kitaev22,schnyderPRB08,schnyderAIP}. Similar to the quantum Hall states, TIs and TSCs are gapped systems hosting gapless modes on their surfaces, insensitive to small perturbations \cite{Hasan2010a,qiZhangRMP11}. However, the robustness of these surface states relies on the existence of discrete antiunitary symmetries, either time-reversal (TRS), $\T$, and (or) particle-hole symmetry, $\C$, being denoted symmetry protected topological (SPT) phases. A natural question is whether SPT phases are robust in the presence of  interactions that do not break explicitly or spontaneously their protecting symmetries. Fidkowski and Kitaev \cite{Fidkowski2010} provided the first example in which an interaction could adiabatically connect two states that are topologically distinct in its absence. This example involved one dimensional (1D) spinless p-wave superconductor with TRS, where it was shown that in the presence of an interaction involving eight Majorana operators the non-interacting topological classification is reduced from $\Z$ to $\Z_8$. Two gapped quadratic Hamiltonians whose topological invariants differ by eight can be connected by a gap-preserving trajectory that involves the interaction.
Consequently, 
 eight Majorana zero modes 
localized at the interface 
between these two phases
will be
gapped by the interaction.
Subsequently, the full topological classification of interacting 1D SPT phases was obtained \cite{Turner2011,Fidkowski2011a,Tang2012}, as well as some examples in two \cite{Yao2013a,Qi2013a,Gu2014} and three dimensions \cite{Wang2014c,Senthil2014a,Vishwanath2013}. An exhaustive classification of 
SPT phases still remains subject of intense research \cite{Senthil2015,Kapustin2014,Gu2012,Bi2015,Wang2015b,Morimoto2015a,You2014a}, where methods such as cobordism, group (super)-cohomology and non-linear sigma model have been recently used.

In this work, we study the reduction of the topological classification of interacting fermionic phases in a given dimension and symmetry class, when this classification in the absence of interactions is $\Z$. That is, we classify interacting fermionic phases which are adiabatically connected to non-interacting ones.
We derive and employ a correspondence between the topological invariant $\nu$ of the $d$-dimensional bulk and the number
$n_{\rm 0D}$ of zero-energy bound states localized at zero-dimensional (0D) topological defects
on its 
surface. 
We argue that when the zero modes, localized in \emph{any} 0D topological defect, are gapped by an interaction,
the state becomes topologically trivial under the same interaction. This constitutes a general criterion on any interaction that allows for a change in the topological sector. 
The argument is made explicit by piercing the surface with a lattice of defects and presenting a concrete example of a quartic interaction that gaps a general surface. Our analysis reproduces the classification obtained in \cite{Morimoto2015a}, without making an assumption on the form of the interaction. Thus, we conclude that lifting the restriction of quartic interactions does not alter the classification.

There is an important distinction between classes with $\Z$ topology in even and odd dimensions. 
The former host {\it chiral} modes on the boundaries whose gapping is forbidden by the conservation of energy; while the latter 
host nonchiral (helical) boundary modes whose protection depends on the presence of chiral symmetry.  Following Refs.~\cite{Qi2013a,Yao2013a}, we 
construct nonchiral SPT phases in even dimensions
by combining two systems with opposite chirality and Chern invariant, adding a $\Z_2$ unitary symmetry $\R$ preventing the coupling of modes with opposite chirality.
The resulting classes, which we refer to as prime ($'$) classes, naturally generalize SPT phases to even dimensions, making our analysis equally applicable to all 
dimensions.

\begin{figure}
\centering
\includegraphics[width=1\columnwidth]{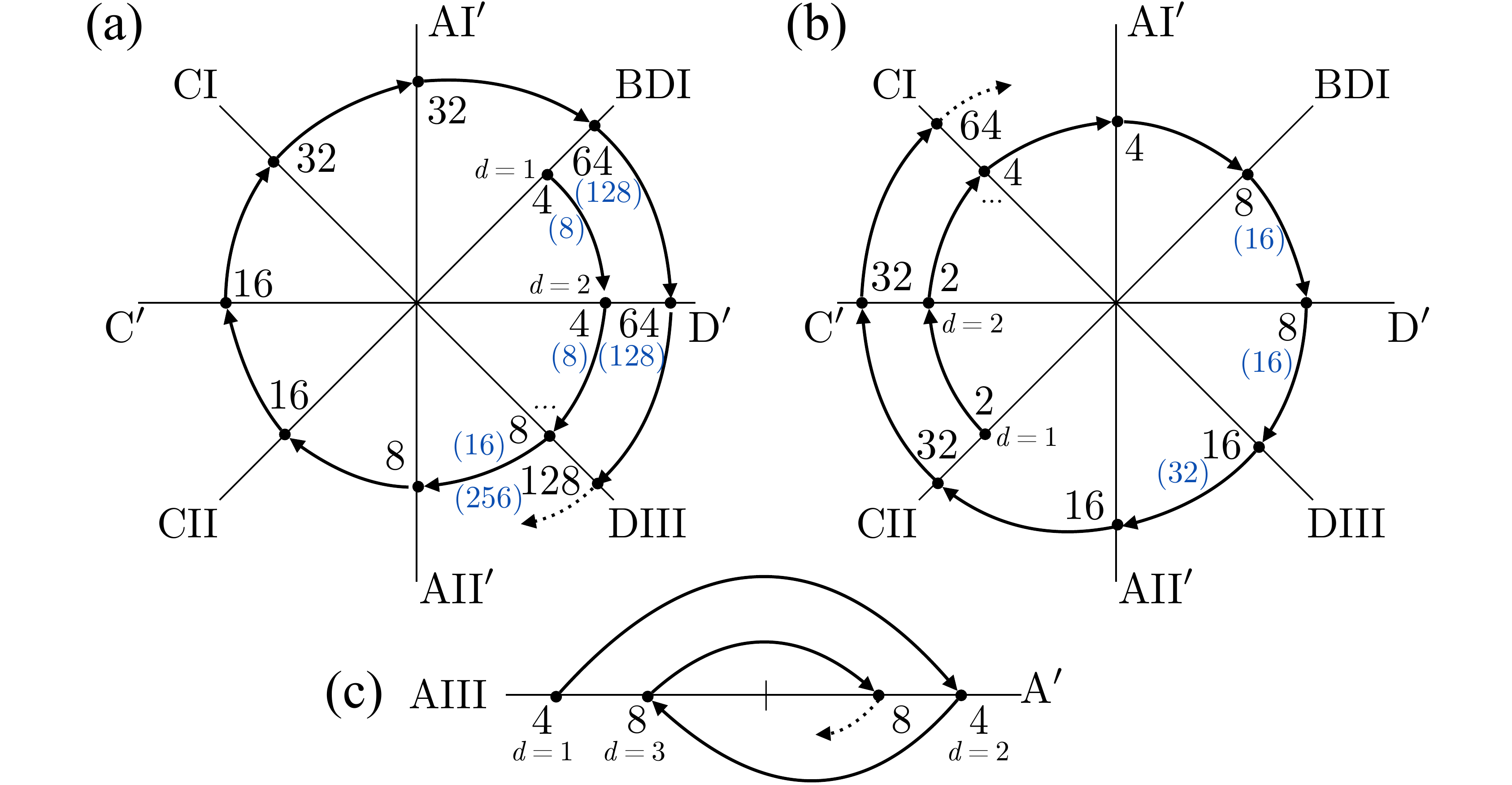}
\caption{
{Classification of interacting fermionic SPT phases with $\Z_n$ topology, with $n$ given by Eq.~\eqref{dimensionalred} (blue for $\mu=2$). The symmetry classes are arranged to form a eight- or two-hour ``{Bott clock}'', starting at the 1D symmetry classes, (a) BDI, (b) CII and (c) AIII, tracing each series clockwise for increasing dimensions. At each even to odd dimension step, $n$ is increased by a factor of 2. The 8(2)-hour periodicity of the real(complex) classes is not satisfied, leading to an infinite spiral.}
 }\label{clock}
\end{figure}

We pursue an analogous path to Ref.~\cite{Teo2010}, in
its description of the ten-fold classification of non-interacting topological states 
using a eight- or two-hour ``Bott clock". We relate the topological character of a SPT class with symmetry $s$ in dimension $d$ 
to one with $s+1$ and $d+1$ (Fig.~\ref{clock}), and find that in
  the presence of interactions, $\Z$ is reduced to $\Z_n$, 
with $n$ violating the clock periodicity.  This is represented as a spiral clock in Figs.\ref{clock}(a)-(c),
distinguished by their $d=1$ symmetry classes:
 (a) BDI (b) CII and (c) AIII. Generally,
\beq
n = 2^{\lfloor \frac{d-1}{2} \rfloor} \mu n_0 \label{dimensionalred},
\eeq
where $n_0=\{4, 2, 4\}$ for (a) to (c), respectively, and $\mu=2$ for classes BDI, D$'$ and DIII, when they are realized by triplet superconductors. For all other cases $\mu=1$.

We start by reviewing how the 0D zero modes at the ends of 1D systems
reflect their topological invariants, and how they may be gapped by interactions~\cite{Tang2012,Fidkowski2011a,Senthil2014a}. We then introduce the 0D surface defects in two and three dimensions, and their reflection of the bulk topological invariant. Following that, we discuss the possible interaction-induced gapping of zero modes within such a defect, the gapping of a lattice of such defects and the derivation of the parameter $n$. Finally, we generalize to any dimension.

\emph{One dimension --}
We focus on Dirac Hamiltonians (DH) as representatives of the different topological sectors ~\cite{ryuNJP2010}. The minimal DH in 
class AIII reads 
\beq
\H_{\rm AIII} = \int \! \d x~ c^{\dagger} (i \sigma^z \partial_x + m(x) \sigma^y) c, \quad c = (c_L , c_R)^T,\label{aiiiham}
\eeq
for $c_{L,R}$ complex fermion operators. Eq.~\eqref{aiiiham} is invariant under the chiral symmetry defined by
$
\S^{-1} c_{L,R} \S = c_{R,L}^{\dagger}$ and $\S^{-1} i \S = -i$.
A 0D edge can be implemented by forcing the mass $m(x)$ to change sign at $x=0$. Choosing $m(x) = \tanh x$ without loss of generality, we obtain the zero energy localized operator  
$\psi = \int \! \d x (c_L + c_R) \sech x$, 
 obeying $\S^{-1} \psi~ \S = \psi^{\dagger}$. For a set of AIII chains there is one zero mode at the end of each chain, so that $n_{\rm 0D}=\nu$, i.e., the number of zero energy end modes equals the topological invariant. Inter-chain coupling is restricted by the symmetry. Mass terms $M_{ij} \psi_i^{\dagger} \psi_j$ are forbidden since hermiticity requires $M_{ij} = M^*_{ji}$, while chiral symmetry requires $M_{ij} = -M^*_{ji}$. Quartic terms, on the other hand, allow for a fully symmetric interaction between the chains
\beq
\H_{\rm int} = V \psi_1^{\dagger} \psi_2 \psi_3^{\dagger} \psi_4 + \text{h.c.}.\label{aiiiint}
\eeq
Here the subscript enumerates the chains.
The interaction (\ref{aiiiint}) has a unique $\S$-symmetric ground state separated by a gap $V$ from the remaining states
given by $\left| 0101 \right> - \left| 1010 \right>$, 
with $0,1$ the 
eigenvalues of 
$\psi^\dagger_i\psi_i$. 
Due to the absence of edge modes, we conclude that a 1D AIII system with  $\nu=4$ becomes topologically trivial once the interaction (\ref{aiiiint}) is included. Thus, $n=4$ for $d=1$ and class AIII.

Analogously, we can describe a chain in class BDI by Eq.~\eqref{aiiiham}, where the fermionic operators $c$ are replaced by Majorana operators $\eta$. 
In this basis, charge conjugation $C$ is the identity, and $\T=\S$.
The edge Majorana bound states are formed by $\gamma = \int \! \d x (\eta_L + \eta_R) \sech x$, localized at $x=0$. Pairing the operators $\gamma_i$  into  complex fermions $\psi_i = \gamma_{2i-1} + i \gamma_{2i}$, 
we see that, Eq.~\eqref{aiiiint} satisfies the symmetries of class BDI and can 
gap out groups of 8 Majoranas, reducing $\Z \rightarrow \Z_8$ ($n=8$).
Similarly, we find $n=2$ in class CII, where the Hamiltonian acts on spinful fermions with an intrinsic double degeneracy.

\emph{From one to two dimensions --}
In 2D we focus on class D$'$. 
The analysis for A$'$ and C$'$ proceeds similarly.
In all cases we find $n_{\rm 0D}=\nu$, such that the step  from $d=1$ to $d=2$ does not increase $n$.
Class D$'$ can be constructed by adding two copies of 2D $p$-wave superconductors with broken TRS, 
bearing electrons of opposite spin direction and gapless edge modes of opposite chirality. The resulting system has TRS  with $\T^2=-1$,
but an additional $\Z_2$ unitary symmetry, $R$, distinguishes it from the class DIII, in which an even number of pairs of counter-propagating edge modes may be gapped without violating TRS~\cite{note}. The symmetry $R$ can be physically implemented either as a conservation of $S_z$ modulo $2$ (conservation of the parity $(-1)^{N_{\up}}$) satisfying $\R^2 = 1$ and $\{\T,\R\}=0$ \cite{Qi2013a,Yao2013a} or as a mirror symmetry satisfying $\R^2 = -1$ and $[\T,\R]=0$ \cite{Yao2013a} in which case D$'$ corresponds to the crystalline phase DIII+R. In both options, the combined operator $\tilde{\T} = \T \R$ is antiunitary and satisfies $\tilde{\T}^2 = 1$. Here, we choose the first $\R$ realization.
The D$'$ system has zero Chern number $N_{\up} + N_{\dn} = 0$ but a non-zero spin Chern number $\frac{1}{2}(N_{\up} - N_{\dn}) $, resulting in a $\Z$ topological classification at the non-interacting level.

The Hamiltonian for $\nu$ pairs of Majorana 1D edge modes is
\beq
\H_{\rm D'} = i \int \! \d x \sum_{a=1}^{\nu} \eta_{a}^{\dagger} \sigma^z \partial_x \eta_{a}, \quad \eta=(\eta_{\up},\eta_{\dn})^T,
\label{HDD}\eeq
with $\T$ and $\R$ defined by $\T^{-1} (\eta_{\up},\eta_{\dn}) \T =  (\eta_{\dn},-\eta_{\up})$ and $\R^{-1} (\eta_{\up},\eta_{\dn}) \R =  (\eta_{\up},-\eta_{\dn})$. 
We relate it to the  BDI Majorana chain studied above by adding
the 
mass term $m(x) \eta^{\dagger} \sigma^y \eta$ at the edge~\cite{Qi2013a}.
Locally, the mass term breaks both $\T$ and $\R$, but preserves the combination $\tilde{\T} = \T \R$. Consequently, the resulting edge Hamiltonian 
has the antiunitary symmetry $\tilde{\T}$ squaring to $+1$, representing spinless fermions (class BDI). { As in the previous section, we choose $m(x)$ to change sign at $x=0$  to form a 0D mass defect and find $\nu$ Majorana bound states
at $x=0$.
Thus, in the noninteracting limit, the 1D BDI system constructed at the boundary inherits the topology of the bulk 2D system of class D$'$. 
If the mass oscillates in space $m(x)=m_0\cos(qx)$, a lattice of 0D defects is formed with a separation of $2\pi/q$. 
When each defect contains eight zero modes, the interaction \eqref{aiiiint} between the zero modes may be used to induce the topological transition from $m_0>0$ to $m_0<0$ without 
closing the energy gap~\cite{Qi2013a} leading to $n=8$.

\begin{figure}
\centering
\includegraphics[width=1\columnwidth]{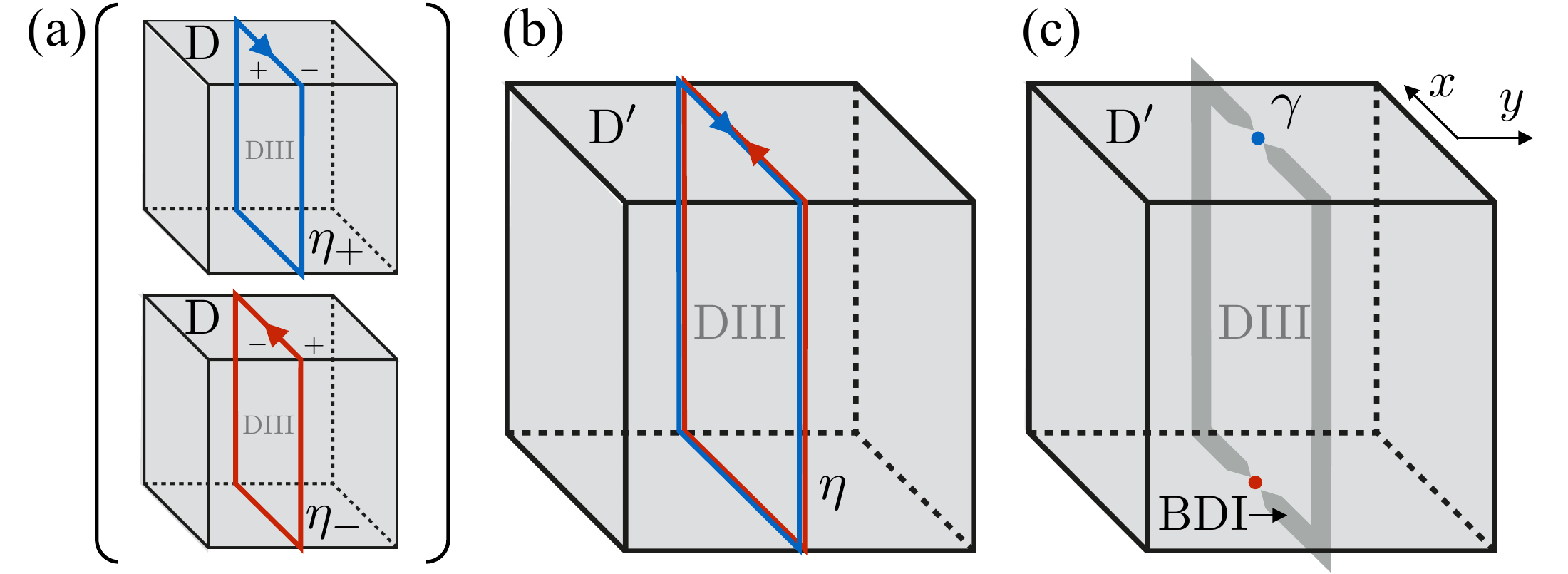}
\caption{ Illustration of the reduction scheme from a 3D DIII system to a 1D BDI system, hosting zero energy modes $\gamma$: (a) Iso-spin multiplet $\chi_a$, Eq.~\eqref{HDIIIm}, of gapped surface modes (light grey) with protected 1D counter-propagating chiral modes $\eta_\pm$ (blue and red lines), localized at $y=0$. (b) Effective D$'$ system with helical mode $\eta$, Eq.~\eqref{HDD}. (c) Gapped 1D BDI system (darker gray) with zero mode (blue and red dots) at $x=y=0$.  }\label{fig1}
\end{figure}

\emph{From two to three dimensions --}
We now follow the same construction to 
 relate class DIII in 3D to its descendant 2D system 
class D$'$ and to its 0D surface defects. In this step we find a doubling of the ratio $\nu/n_{\rm 0D}$, giving rise to a doubling of $n$. Class DIII represents superconductors with $\T^2=-1$.
 At the free fermion level, DIII has a $\Z$ classification where a bulk topological invariant $\nu$ is associated with $\nu$ helical gapless surface modes.
The surface Hamiltonian reads
\beq
\label{HDIII}
\H_{\rm DIII} = i \int \! \d^2 {\bf r} \sum_{a=1}^{\nu} \eta_{a}^{\dagger} [\sigma^z \partial_x + \sigma^x \partial_y] \eta_{a},
\eeq
with $\nu$ taken to be even in the discussion below.
A naive introduction of a surface mass term of the form $m({\bf r}) \eta^{\dagger} \sigma^y \eta$  results in a gapped surface of class D, with $\nu$ {\it chiral} edge modes along a 1D defect (a line along which $m=0$), 
which cannot be gapped. 
Gapping requires two time-reversed copies of class D at the surface, 
obtained by grouping the $\nu$ surface modes into pairs defining an isospin doublet $\chi_a = (\eta_{2a-1}^T,\eta_{2a}^T)$, and adding the mass term $m \chi^{\dagger} \sigma^y \otimes \tau^z \chi$, with  $\tau^i$ the Pauli matrices in isospin space. 
Taking the mass to have an opposite sign 
for the two isospin directions leads to a class D$'$ surface with counter-propagating chiral modes. 
Combining $\T$ with an isospin flip results in a modified TRS (Fig.~\ref{fig1}(a)). The 2D surface Hamiltonian reads,
\beq
\label{HDIIIm}
\H = \int \! \d^2 {\bf r} \sum_{a=1}^{\nu/2} \chi_{a}^{\dagger} [i (\sigma^z \partial_x + \sigma^x \partial_y) \otimes \tau^0 + m({\bf r}) \sigma^y \otimes \tau^z] \chi_{a}.
\eeq
$\H$ is invariant under the modified 
TRS $\T = i \sigma^y \tau^x \K$ and under $\R = \tau^z$ 
satisfying $\T^2 = -1$, $\R^2 = 1$ and  $\{\T,\R\}=0$, 
which implements the D$'$ system 
described above. Choosing $m(\bf r)$ to change sign along the $y$-direction, $m(y) = \tanh y$, we find the 1D gapless modes $\eta_{\pm} = \int \d y \; \sech y ~ v_{\pm} \chi$ with $v_{+} = (1,0,0,0)$ and $v_-= (0,0,0,1)$. The Hamiltonian acting on $\eta_{\pm}$ is given exactly by Eq.(\ref{HDD}) with $\pm$ replacing the spin index.  We can then conclude that on the non-interacting level, the 2D boundary D$'$ system inherits the topology of the 3D DIII system with an invariant $\nu/2$ rather than $\nu$. Adding 
an $x$-dependent mass  $m(x)=\tanh x$, the 1D edge mode becomes gapped and a 0D defect is introduced 
with $\nu/2$ zero modes. 

The surface Hamiltonian (\ref{HDIII}) may be  equivalently written in terms of Dirac matrices, to be 
conveniently generalized to higher dimensions.
We construct the DH
using $\Gamma_{1,\dots,5}$ matrices satisfying the usual Clifford algebra $\{ \Gamma_i, \Gamma_j \} = 2 \delta_{ij}$. The matrices are chosen to be symmetric and antisymmetric for odd and even $i$, respectively. 
The kinetic part of Eq.~\eqref{HDIIIm} becomes $\Gamma_1 \partial_x + \Gamma_3 \partial_y$, and TRS acts as $\T = \Gamma_2 \K$ with $\T^2 = -1$. 
By adding the mass term
$m(x) \Gamma_2 + m(y) \Gamma_4$, we obtain a 0D defect localized at points where both $m(x)$ and $m(y)$ change sign.  The mass term breaks $\T$ of the original Hamiltonian but leads to an emergent antiunitary symmetry given by $\tilde{\T} = \Gamma_5 \K$ with $\tilde{\T}^2 = 1$. That is, the new Hamiltonian is in class BDI, hosting Majorana bound states at the 0D topological defect \cite{Teo2010}.
For  $m(x) = (-1)^{s_x}\tanh x$ and $m(y)=(-1)^{s_y}\tanh y$, the zero modes are given explicitly by $\gamma = \int \! \d^2{\bf r} \; \sech x \; \sech y \; v \chi$, with $v$ the non-zero eigenvector of the projection operator $P = \frac{1}{4} (1 + i(-1)^{s_x} \Gamma_1 \Gamma_2)(1 + i(-1)^{s_y} \Gamma_3 \Gamma_4)$. Here $s_x,s_y$ are integers. The sign changes they introduce emerge naturally when the mass oscillates in space to introduce a lattice of defects{, for example using the mass functions $m(x)=\cos{qx}$ and $m(y)=\cos{qy}$.}

{
A local interaction that renders the system topologically trivial must guarantee that the 0D defect zero modes are gapped by acting in the projected space of zero modes. This is required since the interaction matrix elements which couple the zero and high energy modes can be made arbitrarily small by an appropriate choice of $m$, and the elements coupling zero modes and bulk states vanish by locality. Hence, we conclude that when the defect contains less than 8 zero modes the topological sector is stable to any interaction.
In the case where a 0D defect cannot be constructed, for example for odd values of $\nu$, it is always possible to reduce the 2D gapless modes to a single 1D gapless mode coupled to a number of 0D zero modes. The 1D mode can never be gapped out by interactions, guaranteeing the stability of the topological sector. Put together, these considerations imply the stability of the 2D DIII surface with $\nu < 16$ to any interaction. 

}
For $\nu=16$, 0D defects containing 8 zero modes can be constructed and gapped by interactions. 
We{ can explicitly find one interaction that gaps the full surface by} combining pairs of Majorana surface states into four complex fermions $\psi_i = \chi_{2i-1} + i \chi_{2i}$ in analogy to the 0D case, 
\beq
\H_{\text{int}} = \int \! \d^2 {\bf r} \; V (\psi_1^{\dagger} \Gamma_5 \psi_2) (\psi_3^{\dagger} \Gamma_5 \psi_4),\label{intdiii}
\eeq
with $\Gamma_5=\Gamma_1\Gamma_2\Gamma_3\Gamma_4$. This interaction reduces to the interaction (\ref{aiiiint}) upon projecting to the space of zero modes in the defect. It respects both $\T$ and $\tilde{\T}$ symmetries.

The gapping procedure of the 2D surface is similar to the one employed to gap the 1D surface of the class D$'$ system at $d=2$ \cite{Qi2013a}. In the absence of interaction within each 0D defect, tunnel coupling between zero modes in neighboring defects creates a spectrum that is gapless at zero energy. This spectrum is not identical to the one obtained in the absence of the mass terms, but its low energy characteristics are identical. The zero modes within each defect all share the same values of $s_x,s_y$ and thus the interaction (\ref{intdiii}) respects the symmetry. When the interaction is stronger than the hopping terms between defects, the surface is gapped. This procedure bears similarity to the proliferation of monopoles used in Refs.~\cite{You2014a,You2014d}.

\emph{Higher dimensions ---}
{  We now generalize beyond $d=1,2,3$ to obtain the complete classification of fermionic SPTs , as summarized in Fig.~\ref{clock} and  Eq.~\eqref{dimensionalred}.
  We start with the complex series (Fig.~\ref{clock}(c)), and comment on the extension to real classes. 
 We first identify the correspondence between $\nu$ and $n_{\rm 0D}$, and then  introduce an interaction that gaps the surface once it is pierced with a lattice of 0D defects. The $(d-1)$-dimensional surface 
is described by
\beq
\label{Hdefect}
\H = \int \! \d^{d-1} {\bf r} \; \chi^\dagger\left (i {\boldsymbol \alpha} \cdot {\boldsymbol \nabla} +  {\boldsymbol \beta} \cdot {\bf M}\right )\chi .
\eeq
Here, ${\boldsymbol \alpha} = (\alpha_1, \dots, \alpha_{d-1})$, ${\boldsymbol \beta} = (\beta_1, \dots, \beta_{d-1})$ are Dirac matrices satisfying the Clifford algebra.  
In $d$ spatial dimensions, the minimal massless 
DH with chiral symmetry has dimension $2^{\lfloor \frac{d}{2} \rfloor}$. This can be understood from the doubling of the bulk DH dimension from odd to even $d$, due to the intrinsic doubling of class A$'$, and the fact that the surface and bulk DHs differ in dimension by a factor of two. 

A gapless edge Hamiltonian of a system with a topological invariant $\nu$ may be constructed by $\nu$ copies of the edge  Hamiltonian (\ref{Hdefect}) with $\boldsymbol\beta=0$, enlarging the size of the matrix by a factor $\nu$, giving the combined dimension $\nu 2^{\lfloor \frac{d}{2} \rfloor}$. 
We seek the value $\nu$ for which a mass term that allows the formation of a 0D defect may be introduced.  

A zero mode is an operator that commutes with (\ref{Hdefect}). For a single 0D defect, the $i$-th component of the mass vector 
is chosen to satisfy $M_i=(-1)^{s_i}\tanh r_i$.
The zero mode is then
$\gamma=\int \! \d^{d-1} {\bf r}\prod_{i=1}^{d-1} \!  \sech r_i \; v \chi$, where $v$ corresponds to the non-zero eigenvector of the 
 $d-1$ commuting projection operators
$
P^{s_i}_i = \frac{1}{2} (1 + i(-1)^{s_i} \alpha_i \beta_i)$.
To obtain a single non-zero eigenvalue to this set of operators, we need an Hamiltonian of dimension $2^{d-1}$. The ratio of this dimension to the dimension $\nu 2^{\lfloor \frac{d}{2} \rfloor}$ fixes $n_{\rm 0D}(\nu)$ to be $\lfloor{ {\nu 2^{1-\frac{d}{2}}}}\rfloor$ for even $d$ and $\lfloor{{\nu 2^{\frac{1-d}{2}}}}\rfloor$ for odd $d$.

As before, a group of four complex fermions in a single defect may be gapped by interactions. The surface may be gapped by piercing it by a lattice of 0D defects, each of which containing four zero modes, hence reducing the $\Z$ classification to $\Z_n$, with $n=2^{\frac{d+2}{2}}$ for even $d$ and $n=2^\frac{d+3}{2}$ for odd $d$.

When there are four zero modes in a single 0D defect, one interaction that reduces to Eq.~\eqref{aiiiint} for any choice of $s_i$'s (and therefore for any 0D defect in the lattice)  reads
\beq
\label{Hint}
\H_{\text{int}} = \int \! \d^{d-1} {\bf r} \; V (\psi_1^{\dagger} \Gamma_{2d-1} \psi_2) (\psi_3^{\dagger} \Gamma_{2d-1} \psi_4),
\eeq
where $\Gamma_{2d-1}=\prod_{i=1}^{d-1}\alpha_i\beta_i$.
For values of $\nu<n$, a topologically protected 0D defect cannot be formed, or will host less than four zero modes {that cannot be gapped by any interaction}. Thus, {any interaction involving less that $n$ fermions will be either projected to one that cannot locally gap the surface, or one that is not relevant in the low energy subspace.}

The analogous derivation of $n$ for the real classes finds the same doubling of $n_{\rm 0D}$ 
whenever $d$ is increased by two, as described here for the complex classes. As explained below Eq. (\ref{dimensionalred}), however, the three $\Z$-class series differ in $n_{\rm 0D}$ in the $d=1$ case, and this difference carries over to all larger dimensions. Furthermore, the value of $n$ depends also on the nature of the zero modes being complex or Majorana fermions.

\emph{Conclusion ---} We showed that under interactions the classification of topological phases with chiral symmetry is reduced from $\Z$ to $\Z_n$, and reproduced the value of $n$ derived in \cite{Morimoto2015a}, summarized in Eq.~(\ref{dimensionalred}) and Fig.~\ref{clock}. 
In our approach, the gapless surface separating topologically distinct phases is replaced by a gapless lattice of coupled 0D defects. These defects enclose zero modes, whose number $n_{\rm 0D}$ is determined by dimension, bulk symmetries and bulk topological invariant $\nu$.  We identify the relation between $\nu$ and $n_{\rm 0D}$ and find it to double with an increase of the dimension by two. The number of zero modes for $d=1$ depends on the symmetry of the problem, leading to a difference between the the series $(a)-(c)$ (Fig.~\ref{clock}). 
Our construction establishes the stability of topological phases with $\nu < n$ for any symmetry-preserving interaction, and provides  a necessary and sufficient condition for ones that gap the $n$-boundary modes. It turns out it is always possible to find a quartic interaction satisfying this condition
for an arbitrary dimension and symmetry class, Eq.~(\ref{Hint}).

\acknowledgements{
We acknowledge stimulating discussions with A.P. Schnyder, J.S. Hofmann and B.A. Bernevig. We acknowledge financial support by the European Research Council under the European Unions Seventh Framework Programme (FP7/2007-2013) / ERC Project MUNATOP, Microsoft Station Q, Minerva foundation, and the U.S.-Israel BSF.}

\FloatBarrier

\bibliographystyle{apsrev4-1}

\bibliography{refs}

 \end{document}